\documentclass[twocolumn,prl,showpacs,preprintnumbers,amsmath,amssymb]{revtex4}
\usepackage{amsfonts}

\usepackage{amssymb}
\usepackage{graphics}
\usepackage{epsfig}
\usepackage{epstopdf,graphicx}    
\usepackage{dcolumn}     
\usepackage{amsthm}
\usepackage{bm}          
\usepackage{amsbsy}
\usepackage{color}

\begin{document}

\title{Vacancy Induced Splitting of Dirac Nodal Point in Graphene}

\author{W. Zhu$^{1}$, W. Li$^{1 *}$, Q. W. Shi$^{1}$, X. R. Wang$^{2,3 *}$, X. P. Wang$^{1}$, J. L. Yang$^{1}$, J. G. Hou$^{1}$}
\address{$^1$Hefei National Laboratory for Physical Sciences at Microscale,
University of Science and Technology of China, Hefei 230026, China}
\address{$^2$Department of Physics, The Hong Kong University of
Science and Technology, Clear Water Bay, Kowloon, Hong Kong}
\address{$^3$School of Physics, Shandong University,
Jinan, P. R. China}
\email[Electronic address:]{wliustc@mail.ustc.edu.cn; phxwan@ust.hk}
\date{\today}

\begin{abstract}
We investigate the vacancy effects on quasiparticle
band structure of graphene near the Dirac point.
It is found that each Dirac nodal point splits into two new
nodal points due to the coherent scattering among vacancies.
The splitting energy between the two nodal points is
proportional to the square root of vacancy concentration.
In addition, an extra dispersionless impurity band
of zero energy due to particle-hole symmetry is found.
Our theory offers an excellent explanation to the recent
experiments.
\end{abstract}

\pacs{81.05.Uw, 71.55.-i, 71.23.-k} \maketitle

One of the important features of pristine graphene is the Dirac
nodal structures centered at two inequivalent corners $K$ and
$K'$ of the first Brillouin zone with linear energy dispersion
\cite{Nov,Geim}. The common belief is that the Dirac nodal
structure is robust against the short-ranged potential scattering
because the Fermi wavelength diverges at the Dirac nodal point and
the electronic state correction due to the scatters is negligible
\cite{MacDonald}. However, the recent progresses in the classical
wave physics demonstrates that local resonant structures can
dramatically modify waves whose wavelengths are several orders of
magnitude larger than the structure sizes \cite{Liu,Ebbesen,Moreno}.
Vacancies as well as various chemical adsorbates in graphene
can create resonant states in the vicinity of the Dirac point.
An effect analogous to the classical wave is expected.
Those locally resonant states should dramatically change the
graphene electronic structures and transport properties near the
Dirac nodal point. Indeed, the angle-resolved photoemission
spectroscopy (ARPES) indicates the opening of a tunable band gap
near the Dirac point and the formation of a dispersionless impurity
band in hydrogenated quasi-free-standing graphene
\cite{Vyalikh,Haberer}. Another study of hydrogenated graphene on
SiC showed signals of a metal-to-insulator transition (MIT) due to
the electron localization \cite{Bostwick}. Away from the charge
neutrality point, the transport measurements demonstrated a
sublinear carrier dependence of the conductivity \cite{jhchen}.
Within the Boltzmann transport framework, a sublinear conductivity
in charge density was predicted \cite{Stauber}. Away from the Dirac
nodal point, this theoretical prediction agrees well with
experiments. However, it fails to explain the transport behavior
near the nodal point, which is supposed due to the breakdown of Boltzmann transport
theory there \cite{Robinson,Wehling}. Nevertheless, those predictions have not
included the influence from the substantial change of electronic
structure arising from the resonant scattering. In fact, numerical
simulations show that a dramatic change in the density of states
(DOS) occurs near the nodal point \cite{Pereira,Wu}. Therefore, a
deep understanding of the resonant scattering effects is needed.

In this letter, we present a calculation on the effects of the
vacancy resonant scattering on Dirac nodal structure in graphene.
The quasiparticle dispersion is extracted from the spectral
function $A(\textbf{k},E)$ which can be calculated by extending
the well developed Lanczos approach \cite{Zhu}. In contrast to the previous
theoretical studies of the spectral function by the average
T-matrix approximation (ATA) \cite{Peres,Loktov,Farjam} or the
self-consistent T-matrix approximation (SCTA) \cite{Peres,Tomas},
our proposed method is more general and nonperturbative,
including all the coherent multiple scattering contributions.
The ATA considers only the averaged contribution of impurity
potentials without interference effect from different impurities
while the SCTA includes the partial contribution from quantum
interference.  We found that, instead of only one peak in the
spectral function for a given momentum in the case of a weak
short-range scattering \cite{Zhu}, the resonant scattering yields
multi-peaks in the spectral function. This leads to a complete
change of quasiparticle band structure. Each Dirac nodal point
splits into two new nodal points and the splitting energy between
the two nodal points is proportional to the square root of vacancy
concentration. In addition, an extra dispersionless impurity band
is developed at zero energy. The fact that we can obtain this
existing band proves the accuracy of our method. Our results also
suggest that the Boltzmann theory could work well near the Dirac
point if the modified energy dispersion is taken into account.

$\pi$-electrons of pristine graphene can be modeled by a
tight-binding Hamiltonian on a honeycomb lattice of two sites
per unit cell, $H_0 =t\sum\limits _{<ij>}|i><j|+h.c.$, where
$t$ is the hopping energy between the nearest neighboring atoms.
We consider vacancy disorder potential $V$ in this work because
it gives rise to the resonant scattering which may also arise
from hydrogen or fluorine adsorbates \cite{Ihnatsenka}.
Vacancies are introduced by randomly removing lattice sites
with probability $n_{imp}$ (vacancy concentration).
In the presence of a disorder potential $V$, the single electron
properties can be obtained from the ensemble-averaged Green
function $G(\textbf{k}\pm,E)=\overline{<\textbf{k}\pm|\frac{1}
{E+i\eta-H_0-V}|\textbf{k}\pm>}$, where $|\textbf{k}\pm>$ is
eigenstate of $H_0$.
This Green function can be obtained numerically by using
the Lanczos recursive method \cite{Zhu,Wu,Lanczos,kspace,Wang}.
To numerically obtain an exact ensemble-averaged Green's function
near the Dirac point, a large lattice containing millions sites
($4800\times4800$) is used. The large samples guarantee that the
calculated Green's function is free from the finite size errors.
In the following calculations, the broadening parameter
is set to be $\eta=0.001t$ \cite{Zhu}.

Self-energy function $\Sigma$ is defined in the Dyson's equation as
$G(\textbf{k},E)=G_0(\textbf{k},E)+
G_0(\textbf{k},E)\Sigma(\textbf{k},E)G(\textbf{k},E)$. Thus one has
$\Sigma(\textbf{k},E)=G^{-1}_0 (\textbf{k},E) -G^{-1}(\textbf{k},E)$
\cite{Economou,Simons}. Fig. 1 is the energy-dependence of
calculated real (a) and imaginary (b) parts of the self-energy
function in the conduction band for various momenta $k=0,\ 0.021$
and $0.042$ along $K-M$ direction in the Brillouin zone. To have a
visible effect of the resonant scattering, a relatively large
vacancy concentration $n_{imp}=0.1\%$ is used in the calculation.
Within energy range of $[-0.02t,0.02t]$ (depending on the vacancy
concentration), both $Re\Sigma$ and $Im\Sigma$ depend explicitly on
the momentum, as shown in Fig. 1(c-d). This momentum dependence of
the self-energy results in the failure of the ATA or SCTA
\cite{Montambaux}. It also shows that the effective homogeneous
medium approximation is invalid around the Dirac point. Beyond this
energy range, however, our simulations show insensitiveness of the
self-energy  function to wave vector $\textbf{k}$, and $\Sigma$
varies only with the energy. More interestingly, a spike in
$Re\Sigma$ around the Dirac point is observed in Fig. 1(c), in
contrast with the ATA and SCTA \cite{Peres,Tomas}. Another notable
feature is that the ATA predicts a peak of $Im\Sigma$ at the Dirac
point. Instead, a dip at the Dirac point (Fig. 1(d)) is observed.
Though the self-energy function shows a complicated momentum and
energy dependence around the Dirac point, we find that it still
satisfies the Kramers-Kronig relation \cite{Simons}. In order to
have a better picture of the quasiparticles of the system, we
compute the spectral function below.

\begin{figure}
\includegraphics[width=0.5\textwidth]{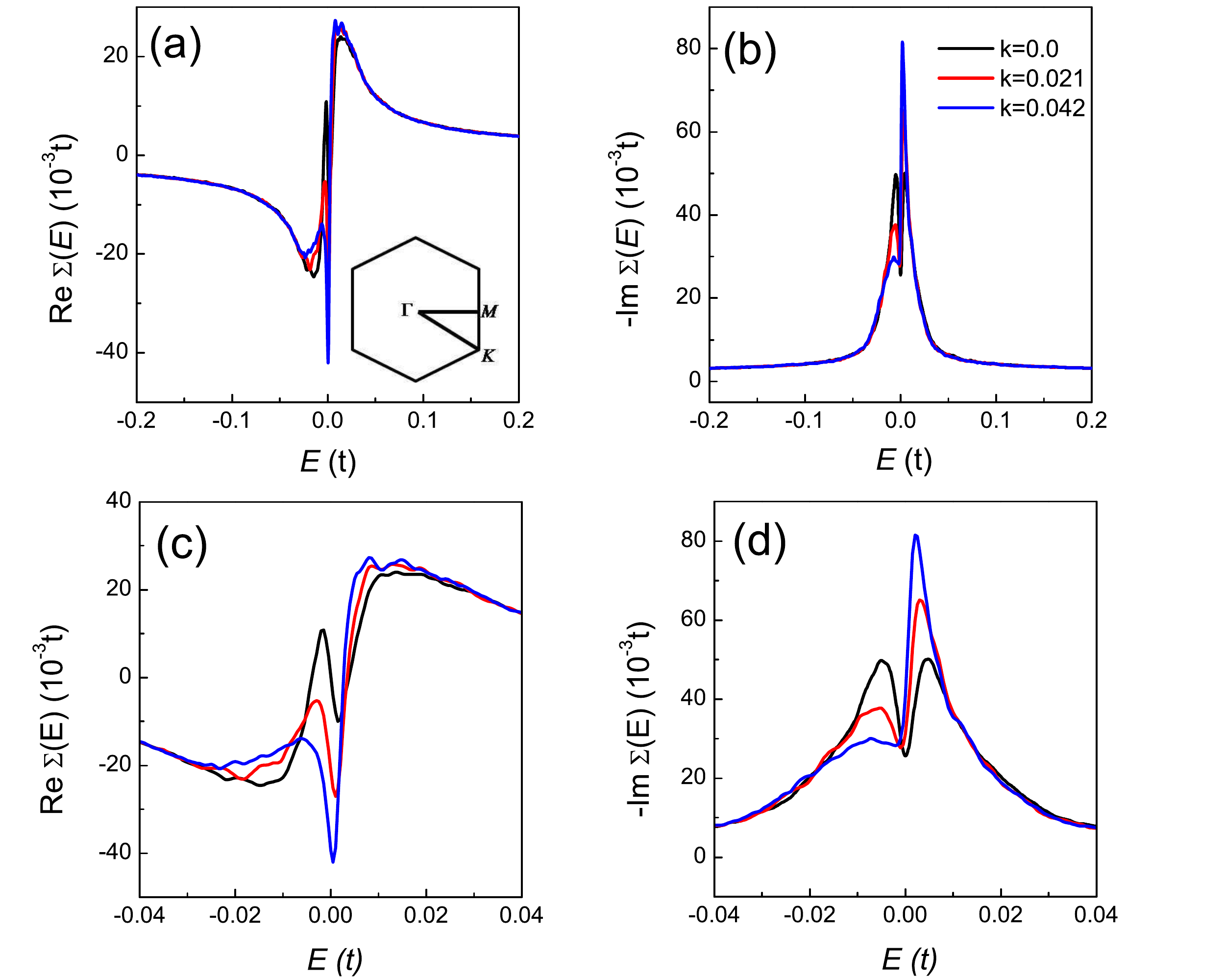}
\caption{(Color online) Real (a) and imaginary (b) parts of
self-energy function $\Sigma$ as a function of energy $E$ for
various wave vectors: $k=0$ (black), $0.021$ (red) and $0.042$
(blue) (in unit of $1/a$). The wave vector varies from the
$K$ point to $M$ point in the Brillouin zone (Inset of (a)).
(c) and (d) are the enlarged views of (a) and (b) near the Dirac
point ($E=0$).} \label{selfE}
\end{figure}

The single-particle spectral function relates to the Green's
function through $A(\textbf{k}\pm,E)=-ImG(\textbf{k}\pm,E)/\pi$
\cite{Simons}. Generally speaking, the spectral function
$A_0(\textbf{k},E)$ is a $\delta$-function in the absence of
disorders, reflecting that the wave vector $\textbf{k}$ is a good
quantum number and all its weight ratio is precisely at energy
$E=E_{k\pm}$. In the presence of disorders, the $\delta$-peak is
broadened due to finite life time of quasiparticles, resulting from
the disorder scattering effect. The linewidth of the peak is given
by $Im\Sigma(E)$ that measures the elastic relaxation lifetime
$\tau_e$, $\tau_e=\frac{\hbar}{-2Im\Sigma(E)}$. These general
features are indeed observed in the weak scattering cases
\cite{Zhu}. The spectral function is qualitatively different in the
strongly resonant scattering regime. Taking $A(k=0{+},E)$ in Fig.
2(a) as an example, the spectral function is surprisingly split into
three peaks: One broadened peak ($p_{-}$) centering in the hole
regime ($E<0$), one broadened peak ($p_{+}$) centering in the
electron regime ($E>0$) and one sharp peak ($p_{0}$) at charge
neutrality point ($E=0$). The $p_{-}$ peak moves toward $E=0$ as the
wave vector $\textbf{k}$ increases while its height is reduced and
its width increases. $p_{0}$ peak position do not change with
$\textbf{k}$ while its height decreases with $\textbf{k}$.
Meanwhile, the $p_{+}$ peak moves away from $E=0$ and its height is
significantly increased and its width is narrowed. When the wave
vector exceeds a threshold value, the $p_{-}$ and the $p_0$ peaks
disappear.

By tracing the trajectories of the peaks of the density plot of
spectral function $A(\textbf{k},E)=A(\textbf{k}+,E)+A(\textbf{k}
-,E)$ in the $k-E$ plane, shown in Fig. 2(b), one can obtain the
dispersion relation $E(\textbf{k})$ shown in Fig. 2(c). Many other
physical quantities like the group velocity and elastic scattering
time can be directly extracted from Fig. 2(b-c). Away from the Dirac
point, the $p_{+}$ peak dominates the spectral function and the
effective energy dispersion approaches the linear behavior (red
square dotted line). Around the Dirac point, Fig. 2(b) clearly shows
the anomalous band structure of quasiparticle. The dispersionless
impurity band (green diamond dotted line) corresponding to the $p_0$
peak in the spectral function indicates the localized states
relating to the kink in $Re\Sigma$ and the dip in $Im\Sigma$. The
existence of these localized states are originated and protected by
the particle-hole symmetry \cite{Castro}: Each vacancy deducts one level from the
continue band, and at the same time pulls out another level from the
continue band to a highly localized zero energy state (hopping is
not allowed in order to be dispersionless) \cite{yang}. Using the
exact diagonalization method, one can also see from the inverse
participation ratio analysis that these impurity states are highly
localized around each vacancy \cite{Pereira}. Applying the Stoner
theory for magnetism, this zero-bandwidth dispersionless band may
lead to ferromagnetism for the system when the electron-electron
interaction is switched on. This observation is consistent with
the recent experiment that directly confirms the dispersionless
impurity band in graphene \cite{Haberer}. More importantly, a new
dispersion quasiparticle band (blue circles) with finite lifetime
corresponding to the $p_{-}$ peak in the spectral function is
observed in our calculations. This indicates that a splitting
of host band of pristine graphene happens near the Dirac point.
\begin{figure}
\includegraphics[width=0.5\textwidth]{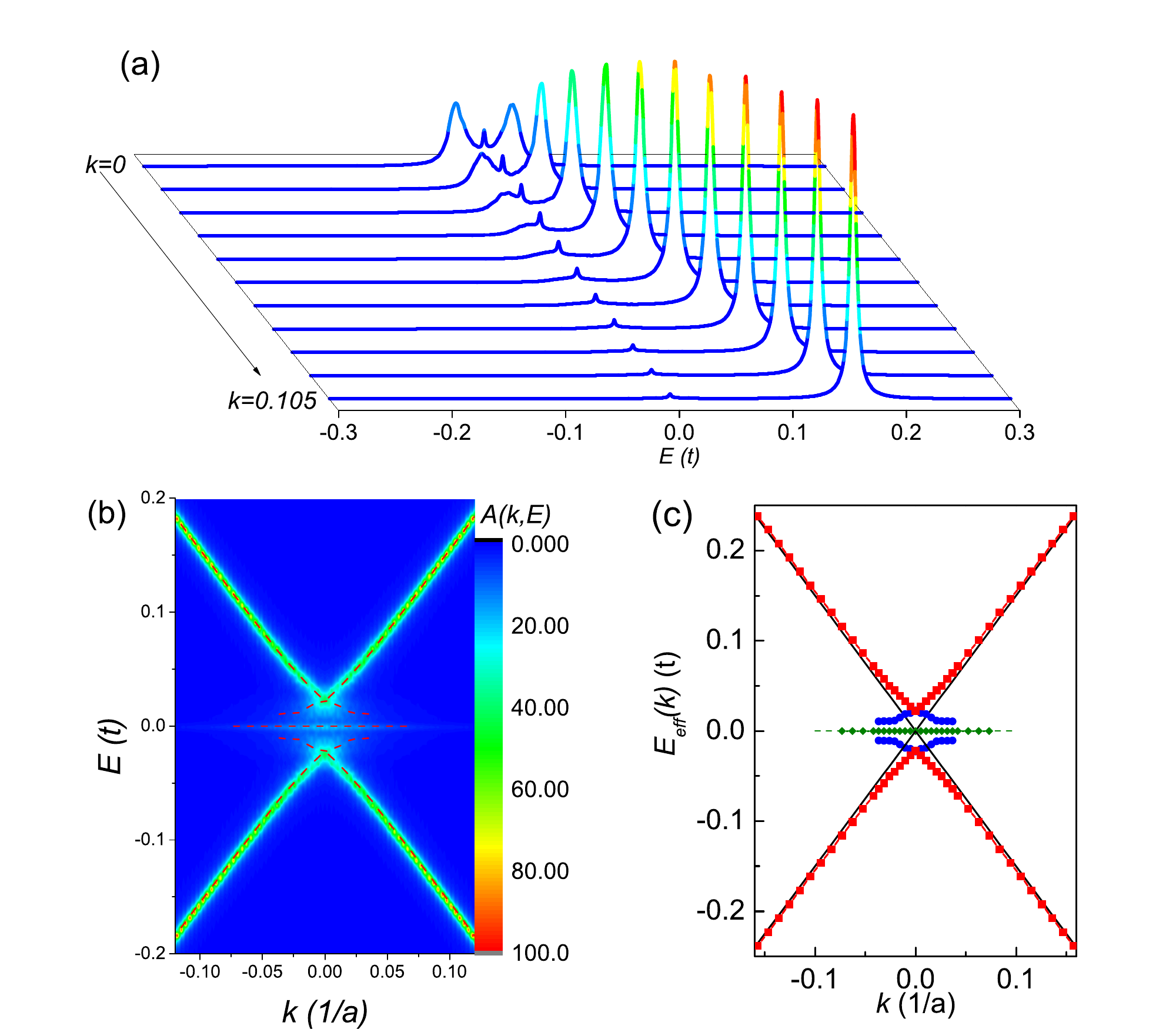}
\caption{(Color online) (a) Quasiparticle spectral function
$A(\textbf{k}+,E)$ plotted as a function of energy $E$ for
various wave vectors at vacancy concentration $n_{imp}=0.1\%$.
$A(\textbf{k}-,E)$ can be obtained by reflecting $A(\textbf
{k}+,E)$ around $E=0$ due to the particle-hole symmetry.
(b) Density plot of spectral function $A(\textbf{k},E)=A(\textbf{k}
+,E)+A(\textbf{k}-,E)$ in $\textbf{k}-E$ plane for $n_{imp}=0.1\%$.
(c) Dispersion relation (dotted line) extracted from (b).
The blue circles and green diamonds are the new resonant band
and dispersionless impurity band, respectively. For comparison, the
linear dispersion of clean graphene is plotted as the black solid
lines.}\label{ak}
\end{figure}

The physical origin of above results comes from vacancy effect.
Firstly, they create zero energy dispersionless impurity band,
forming resonance scattering centers. The strong resonance
scattering around the Dirac point where the electronic de Broglie
wavelength $\lambda$ is much bigger than the average
vacancy-to-vacancy distance $L_v\sim1/\sqrt{n_{imp}}$. The coherent
scattering between the neighboring vacancies mix $K$ and $K'$ points
\cite{Zhangyy} so that a band gap is opened at the Dirac point. This
coupling between the resonant scatters results in the appearance of $p_{-}$ and $p_{+}$
peaks for both $A(\textbf{k}+,E)$ and $A(\textbf{k}-,E)$. In fact,
the direct evidence of the corresponding splitting of spectral
function was indeed reported in hydrogenated graphene by ARPES
\cite{Vyalikh,Haberer}. As the wave vector exceeds a threshold so
that the wavelength is small $\lambda\ll L_v$,  the quasiparticle is
insensitive to the coupling of resonant states between neighboring
vacancies in the low
density. Hence, the new dispersive band disappears for the short
wavelength. This is why ATA approximate approach works well far
enough from the Dirac point.

\begin{figure}
\includegraphics[width=0.45\textwidth]{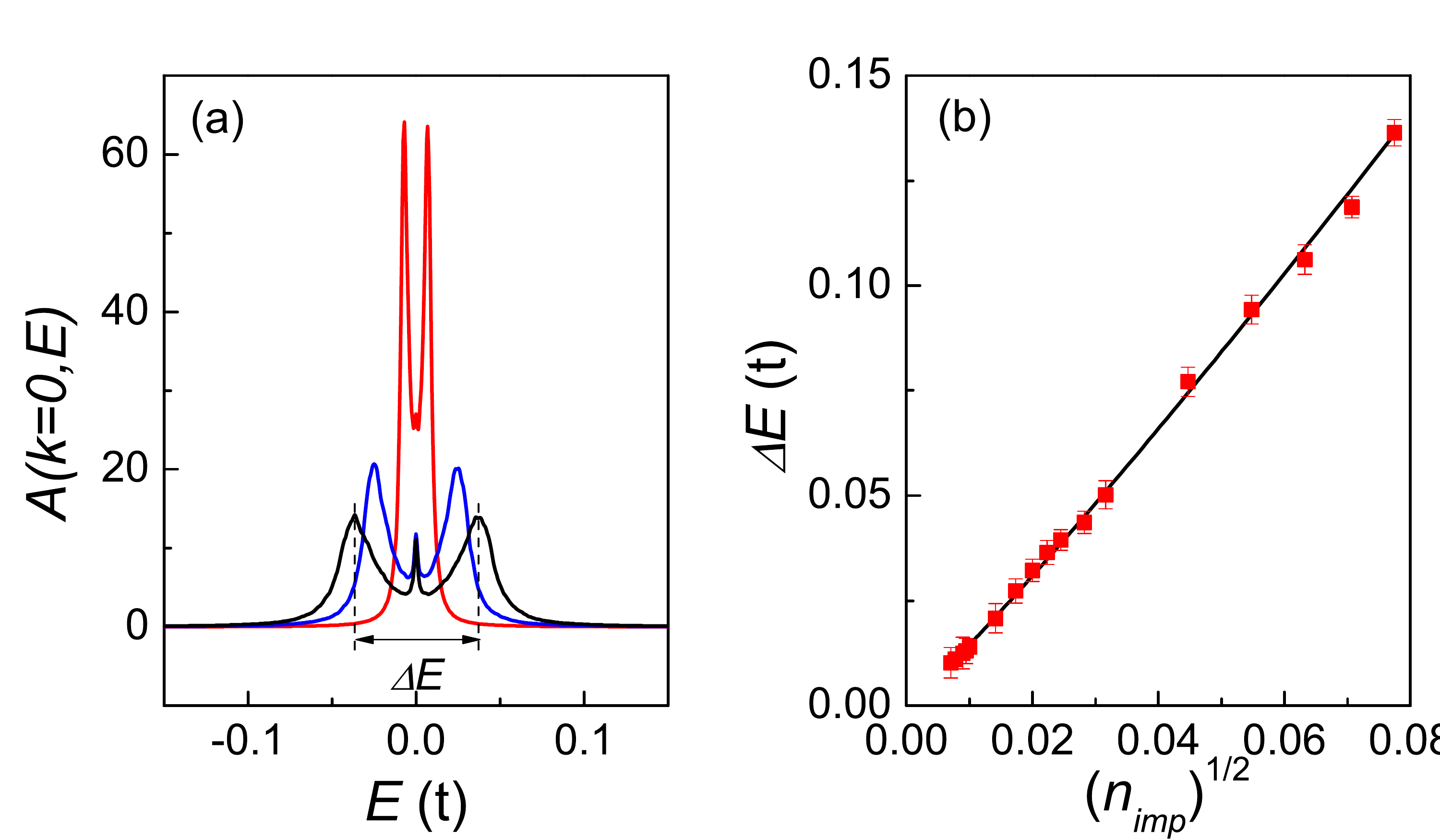}
\caption{(Color online) (a) Quasiparticle spectral function
$A(\textbf{k}=0,E)$ as a function of energy $E$ for several vacancy
concentrations: $n_{imp}=0.01\%$ (red), $n_{imp}=0.1\%$ (blue), and
$n_{imp}=0.2\%$ (black). $\Delta E$ is the splitting energy between
the two resonant peaks. (b) The vacancy concentration dependence of
the splitting energy $\Delta E$ (red dots). The x-axis is in
$(n_{imp})^{1/2}$. The black line is the linear fit $\Delta
E=A(n_{imp})^{1/2}$ with $A=1.76$.}\label{an}
\end{figure}

Coming back to the spectral function at wave vector $\textbf{k}=0$
(as shown in Fig. 3(a)), one striking feature is the splitting
energy $\Delta E$ (peak-to-peak distance between $p_{+}$ and $p_{-}$
peaks). $\Delta E$ is the splitting energy between two nodal points
at $\textbf{k}=0$ (neglecting the impurity band). The vacancy
concentration dependence of $\Delta E$ is plotted in Fig. 3(b),
where the x-axis is in the square root of $n_{imp}$ that measures
the inverse of vacancy-to-vacancy distance. The square root
dependence of $\Delta E\propto \sqrt{n_{imp}} \propto 1/L_v$
supports the picture that the coherent scattering between vacancies
play an important role.

The qualitative changes of the original linear dispersion relation
around the Dirac point give rise to a completely different DOS near
the zero energy. The DOS $\rho$ can be directly computed from the
spectral function by $\rho(E)=\frac{1}{N} \sum_{k}A(k,E)$. As shown
in Fig. 4(a), a shape singular peak due to the dispersionless
impurity band appears at $E=0$. The smooth part of DOS has also a
peak near $E=0$ that is from the two extra dispersion relations
(blue circles in Fig. 2c). The DOS far away from $E=0$ are
essentially from the original linear dispersion relation of pristine
graphene. The width of the energy regime influenced by vacancies is
approximately proportional to $\Delta E\sim \sqrt{n_{imp}}$. All of
these results agree exactly with the the previous numerical
simulation \cite{Wu,Pereira} and recent APRES experiment
\cite{Haberer}. Thus, it proves that our spectral analysis captures
the correct physics near the Dirac point. Furthermore, it shows that
ATA or SCTA fail to describe this interesting and important feature
of DOS \cite{Peres}.

\begin{figure}[t]
\includegraphics[width=0.5\textwidth]{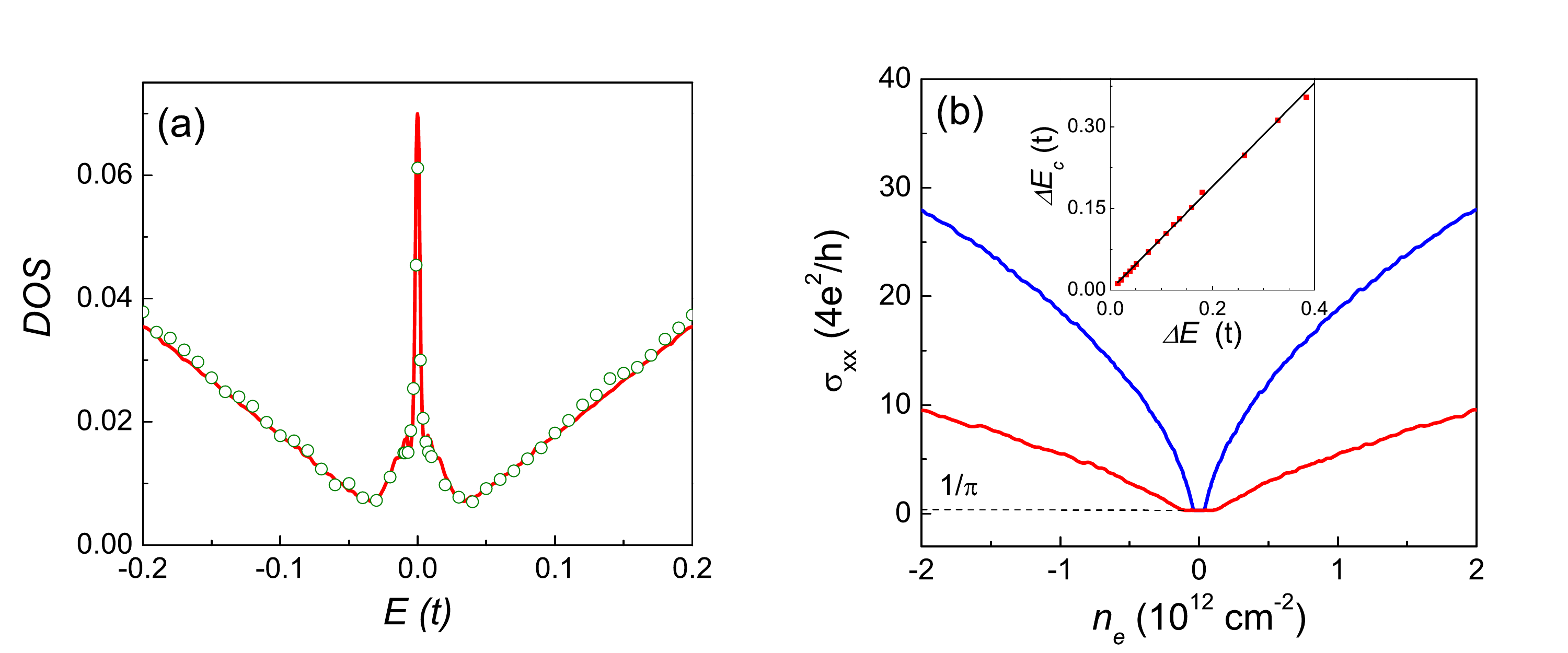}
\caption{(Color online) (a) Comparison of the DOS obtained from
our spectral function (red line) with the numerical simulation
of reference \cite{Wu} (green dots) for $n_{imp}=0.1\%$.
Excellent agreement demonstrates the accuracy of our calculations.
(b) Conductivity as a function of charge density $n_e$ for
$n_{imp}=0.01\%$ (blue) and $n_{imp}=0.1\%$ (red), respectively.
Inset: The conductivity plateau width in terms of energy $\Delta E_c$ vs splitting energy $\Delta E$ shown in Fig. \ref{an}.
$\Delta E_c$ can be obtained from the relationship $n_e=\int^{E_F}\rho(E)dE$. Black line is the
linear fit of $\Delta E_c=A \Delta E$ with $A=0.95$.}\label{dc}
\end{figure}

The new feature of DOS near the zero energy has important
consequences in electron transport. To clearly see it, we calculate
the conductivity by using the Kubo formula at zero temperature,
\begin{eqnarray*}
\sigma_{xx}=\frac{\hbar}{2\pi L^2}\overline{Tr[j_x G^R(E) j_x
G^A(E)]}.
\end{eqnarray*}
Neglect the corrections to current vertex and use the
Drude-Boltzmann approximation
$\overline{G^RG^A}\simeq\overline{G^R}\ \overline{G^A}$
\cite{Montambaux}, our calculated conductivity as a function of
charge density for $n_{imp}=0.01\%$ and $0.1\%$ is plotted in Fig.
4(b). A plateau of conductivity $4e^2/\pi h$ is observed near the
charge neutrality point. The width of the plateau increases with the
vacancy concentration. Away from the plateau, the conductivity
increases with charge density sublinearly. These results are
consistent with those of the numerical real space calculations
\cite{Wehling,Yuan,Ferreira}. Interestingly, plateau width in terms
of energy $\Delta E_c$ is almost equal to the splitting energy
$\Delta E$ as shown in the inset of Fig. 4(b), which
implies that the origin of the plateau is the anomalous band
structure near the Dirac point. This conductivity plateau provides
an interpretation to the experimental observation \cite{Ni}.
It should be pointed out that our calculations do not include
the contribution from the impurity-band-induced variable range
hopping conduction which may also be important in understanding
the experiments at finite temperatures \cite{jhchen,Haberer}.

In conclusion, we have studied vacancy induced resonant scattering
effects on the one-electron properties of graphene. From our
accurate spectral function, a new quasiparticle dispersion band due
to the coherent scattering between neighbor vacancies is predicted,
and each Dirac point splits into two new nodal points. Furthermore,
a dispersionless impurity band is developed at zero energy. This
result sheds light on the characteristics of electronic structures
and transport properties of graphene in the presence of resonant
impurities.

This work is partially supported by NNSF of China (Nos.
10974187,10874165), by NKBRP of China (No. 2011CB921403, No. 2012CB922003),
and by KIP of the CAS (No. KJCX2-YW-W22).
XRW acknowledges the support of Hong
Kong RGC grants (\#604109, RPC11SC05, and HKUST17/CRF/08).

\end{document}